\documentclass[twocolumn,showpacs,preprintnumbers,amsmath,amssymb,superscriptaddress,aps,prb]{revtex4-1}

\usepackage{epsfig}

\usepackage{graphicx,latexsym,xcolor}

\usepackage{tikz}

\usepackage[english]{babel}
\usepackage{amsmath}
\usepackage{mathtools}
\usepackage{color}
\usepackage{epsfig}
\usepackage{graphicx}
\usepackage{bm}
\usepackage{times,amsmath,amssymb}
\usepackage{subfigure}
\usepackage{amsfonts}
\usepackage{amssymb}
\usepackage{color}
\usepackage{slantsc}
\usepackage{array}
\usepackage[T1]{fontenc}

\global\long\def\bege{\begin{equation}}
\global\long\def\ende{\end{equation}}

\global\long\def\begal{\begin{align}}
\global\long\def\endal{\end{align}}

\begin{document}


\title{Robust non-integer conductance in disordered 2D Dirac semimetals \\}

\author{Ilias Amanatidis}
\affiliation{Department of Physics, Ben-Gurion University of the Negev, Beer-Sheva 84105, Israel }

\author{Ioannis Kleftogiannis}
\affiliation{Physics Division, National Center for Theoretical Sciences, Hsinchu 30013, Taiwan}


\date{\today}
\begin{abstract}
We study the conductance $G$ of 2D Dirac semimetal nanowires at the presence of disorder. For an even nanowire length $L$ determined by the number of unit cells, we find non-integer values for $G$ that are independent of $L$ and persist with weak disorder, indicated by the vanishing fluctuations of $G$. The effect is created by a combination of the scattering effects at the contacts(interface) between the leads and the nanowire, an energy gap present in the nanowire for even $L$ and the topological properties of the 2D Dirac semimetals. Unlike conventional materials the reduced $G$ due to the scattering at the interface, is stabilized at non-integer values inside the nanowire, leading to a topological phase for weak disorder. For strong disorder the system leaves the topological  phase and the fluctuations of $G$ are increased as the system undergoes a transition/crossover toward the Anderson localized(insulating) phase, via a non-standard disordered phase. We study the scaling and the statistics of $G$ at these phases. In addition we have found that the effect of robust non-integer $G$ disappears for odd $L$, which results in integer $G$, determined by the number of open channels in the nanowire, due to resonant scattering. 
\end{abstract}

\maketitle

\section{Introduction}
Quantum transport in topological matter at the presence of impurities (disorder) can reveal various
interesting phenomena, such as robust conductance and phase transitions between topological and other phases of matter occurring in disordered mesoscopic systems such as the Anderson localized (insulating) phase. Topological materials usually contain Dirac cones/points in their band structure which are related to their topological properties. A well known example is graphene, a 2D layer of carbon atoms arranged in a honeycomb lattice structure, where the addition of spin-orbit-coupling(SOC) opens a topological energy gap at the Dirac points and creates topological edge states therein\cite{kane}. Other 3D materials such as the Dirac semimetal, host Dirac cones/points, that are topologically protected by inversion and time reversal symmetries resulting in various diverse topological phases\cite{pixley,roy,borishenko,liu}. Extensive theoretical and experimental studies have been performed for such types of materials. In general, Dirac materials with topological properties such as the $Cd_{3}As_{2}$ and $Na_{3}B$ can be experimentally fabricated with various methods \cite{borishenko,liu}. 

Recently a new class of 2D topological materials has been introduced which host Dirac points protected by non-symmorphic symmetries\cite{young}. These materials dubbed 2D Dirac semimetals are special compared to graphene and its 3D counterparts, because the presence of SOC does not result in a topological energy gap at the high symmetry points. By tuning properly the non-symmorphic symmetries an interchange between a novel topological phase and the standard insulating phase occurs\cite{young}. Additional terms in the  Hamiltonian of these materials, that break the time-reversal symmetry can lead to the 2D analog of 3D Weyl semimetals which is accompanied also by edge states\cite{meidan}. In addition the 2D Dirac semimetals are more tunable than their 3D counterparts, based on theoretical proposals via first principles calculations\cite{Bo,chen,Zhong,wang1,wang2}. Some examples are the following monolayer materials, $Be_{2}C$, $BeH_{2}$ and the KHgSb\cite{ma} while similar types of materials have also been experimentally studied by angle-resolved photoemission spectroscopy(ARPES) experiments\cite{fong}.

 Impurities (disorder) are an inevitable factor in most mesoscopic systems studied in condensed matter physics experiments. Moreover disorder can be used as a theoretical tool to study transitions between different phases of matter like, the metallic, the Anderson localized (insulating) and other ones, topological in nature, such as those occurring in topological materials\cite{efetov1,efetov2,kleftogiannis}. For example, it has been shown that the presence of short-range disorder in the bulk of 3D Dirac materials such as the Weyl semimetal can lead to three distinct phases\cite{pixley,roy}, incompressible semimetal, metal and Anderson insulator as the disorder strength is varied. In addition the wavefuctions of the electrons reveal multifractal properties at the critical points of the phase transitions.

In this paper we study the quantum transport properties of 2D Dirac semimetal nanowires, at the presence of disorder. The study is carried out via the Green's function technique based on the Landauer formalism for the conductance at the Dirac point of the 2D Dirac semimetals. For even nanowire lengths $L$, determined by number of unit cells, we find non-integer values for the conductance $G$ through the nanowire, that are robust to the disorder and $L$. The effect is created by a combination of the scattering effects at the contacts(interface) between the leads and the nanowire(scatterer), an energy gap present for the nanowire with even $L$ and the topology of the semimetals, which stabilizes the non-integer $G$ inside the nanowire. The robust non-integer values of $G$ are controllable via the width of the nanowire and the material used in the leads. As the disorder strength is increased a transition towards the Anderson localized (insulating) phase occurs. We study the statistics and the scaling of $G$ as the system crosses over to the insulating phase. The nanowire for odd $L$ gives integer values of $G$ as in other topological systems, such as the Kane-Mele(KM) model at the presence of disorder. The integer $G$ in this case is determined by a resonant scattering mechanism due to the existence of persistent states at the Dirac point where the quantum transport is studied.

\section{2D Dirac semimetal and quantum transport modelling}
\begin{figure}
\begin{center}
\includegraphics[width=0.8\columnwidth,clip=true]{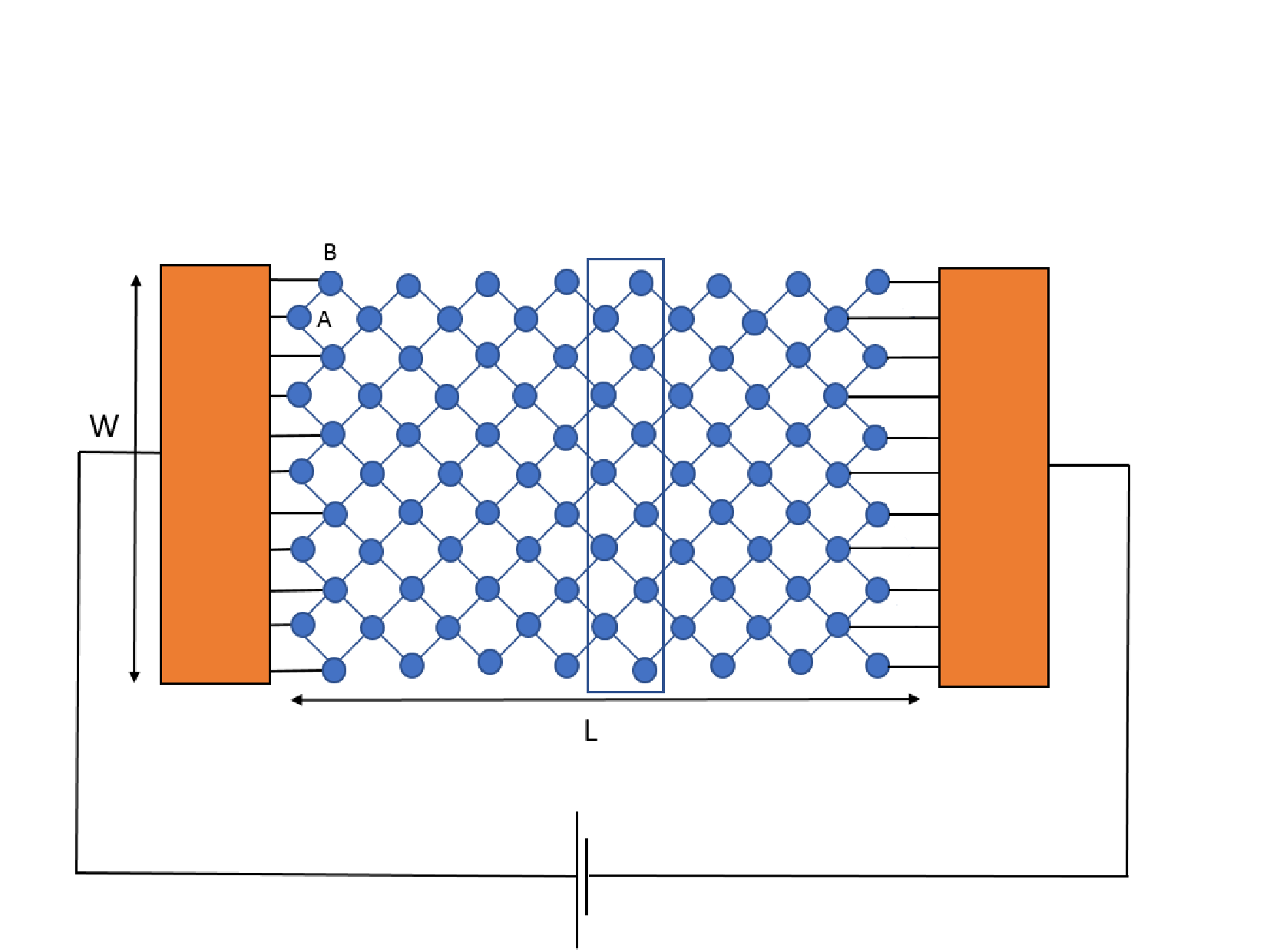}
\end{center}
\caption{Schematic of the studied system, consisting of  the 2D Dirac semimetal nanowire(blue tilted square lattice) of length $L$ and width $W$, sandwiched between two metallic leads (orange stripes). The 2D Dirac semimetal is modelled by a tilted tight-binding square lattice with two types of atoms A and B and spin-orbit coupling (SOC) between them. The unit cell consisting of a zig-zag chain is shown inside the box.
The leads are modeled by a square tight-binding lattice or single chains.}
\label{fig1}
\end{figure}

In order to study the conductance of the 2D Dirac semimetal, we use the  Landauer approach, via the recursive  Green's function technique. In the Landauer approach, the system is build by attaching a left and a right metallic lead at the two ends of the nanowire, which acts as a scatterer and whose conductance we want to calculate. A schematic of this setup is shown in figure \ref{fig1} where the length $L$ is measured in number of unit cells across the nanowire, while the width $W$ is measured in number of sites inside each unit cell, transverse to the transport direction. We model the leads by a square tight-binding lattice described by the Hamiltonian 
\begin{equation}
H_{leads} =   t_{l}\sum_{\substack{ \text{$<${\fontfamily{cmr}\selectfont\itshape i,j}$>$ }\\ \text{a= }\uparrow, \downarrow}} c_{ia}^{\dagger}c_{ja}, 
\label{eq_1}
\end{equation}
where $c_{ia}^{\dagger},c_{ia}$ are the  creation and annihilation operators for spin $a=\uparrow, \downarrow$ at site $i$ and brackets $<$ $>$ denote nearest neighboring sites in the lattice. In our calculations, we consider the value of the hopping $t_{l}=1eV$.

The 2D Dirac semimetal nanowire with non-symmorphic symmetries can be modeled by a tilted tight-binding square lattice with a spin-orbit coupling (SOC) interaction between next-nearest neighboring atoms(see figure \ref{fig1})\cite{young,meidan}. Equivalently if the titled square lattice is split in two sublattices A and B then the SOC interaction connects sites that belong to the same sublattice. In addition we include an on-diagonal(site) random potential to simulate the short-range disorder. The Hamiltonian of the system in second quantization form is given by
\begin{equation}
\begin{split}
H & =  \sum_{\substack{ i \\ \text{a=}\uparrow, \downarrow}} \epsilon_{ia}c_{ia}^{\dagger}c_{ia}+   t\sum_{\substack{ \text{$<${\fontfamily{cmr}\selectfont\itshape i,j}$>$ }\\ \text{a,b= }\uparrow, \downarrow}} c_{ia}^{\dagger}c_{jb} +  \\
& \text{i} t_{so}\sum_{\substack{\ll \text{{\fontfamily{cmr}\selectfont\itshape i,j} }\gg\\ \text{a,b= }\uparrow, \downarrow}}\nu_{ij}c_{ia}^{\dagger}(\boldsymbol{\sigma} \times
\widehat{d}_{ij})^{z}c_{jb},
\end{split}
\label{eq_2}
\end{equation}
where $\ll$  $\gg$ indicate next-nearest neighboring sites,  $\widehat{d}_{ij}$ is the unit vector that connects sites $i$ and $j$ and $\boldsymbol \sigma$ denotes the vector of Pauli matrices. The parameter $\nu_{ij}$ is an integer which takes values +1 when it connects atoms of type B and -1 when it connects atoms of the type A or vice versa.
Parameter $t_{so}$ is a next-nearest neighbor hopping amplitude measuring the spin-orbit interaction strength.
In our calculations we set $t=t_{l}=1eV$. Finally the parameter $\epsilon_{ia}$ is a random potential at site $i$ sampled from a uniform distribution in the interval $[-w/2,w/2]$ with $w$ denoting the strength of the disorder. 
In order to calculate the conductance through the 2D Dirac semimetal nanowire, we use the Landauer formalism based on the recursive Green's function method\cite{li,lewenkopf,paper1,paper2}. The conductance $G$ is expressed in units of the conductance quantum $e^{2}/h$, while the energy is expressed in units of the hopping $t$. We study $G$ at the Dirac point of the 2D Dirac semimetal at nominal energy $E=10^{-7}$ which is the Fermi energy of the system.

\begin{figure}
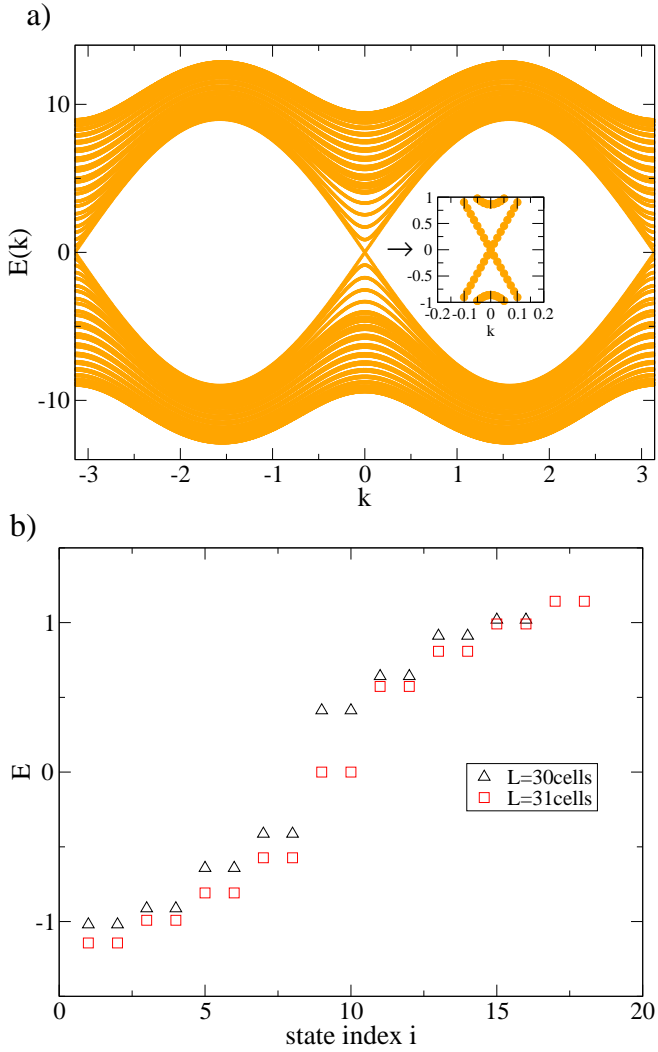

\begin{center}
\includegraphics[width=1.0\columnwidth,clip=true]{fig2a.eps}
\includegraphics[width=1.0\columnwidth,clip=true]{fig2b.eps}
\end{center}
\caption{ a) The bandstructure of the 2D Dirac semimetal nanorire of width $W=61$ with spin-orbit coupling (SOC) strength $t_{so}=4.5eV$. Inset: A linear energy dispersion is followed near the Fermi energy $E=0$. b) The energy levels of the isolated nanowire without the leads for width $W=61$, for two different lengths, $L=30$(open black triangles) and $L=31$(open red squares). At $E=0$ there are two states for $L=31$ which are absent for $L=30$ which has an energy gap.}
\label{fig2}
\end{figure}

\begin{figure}
\begin{center}
\includegraphics[width=1.0\columnwidth,clip=true]{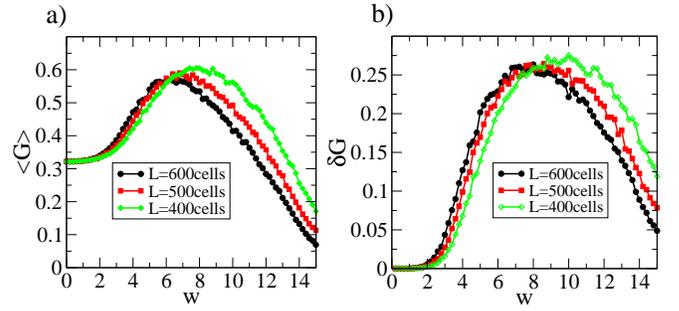}
\end{center}
\caption{The conductance statistics in the presence of on-diagonal short-range disorder for a system with width $W=61$, with even lengths $L$ and spin-orbit coupling (SOC) strength $t_{so}$=4.5eV, for 10000 runs. a) The mean value of the conductance $\langle G \rangle$ and b) its fluctuations(variance) $\delta G$ vs the strength of disorder $w$. The system undergoes a transition from a topological phase with robust non-integer $G$ for weak disorder $w \lesssim 2$ with $\delta G=0$, to a non-standard disordered phase for $ 2\lesssim w \lesssim 10 $, indicated by the increase of $\langle G \rangle$, and finally to the Anderson localized (insulating) phase for strong disorder $w>10$.}
\label{fig3}
\end{figure}

\begin{figure}
\begin{center}
\includegraphics[width=1.0\columnwidth,clip=true]{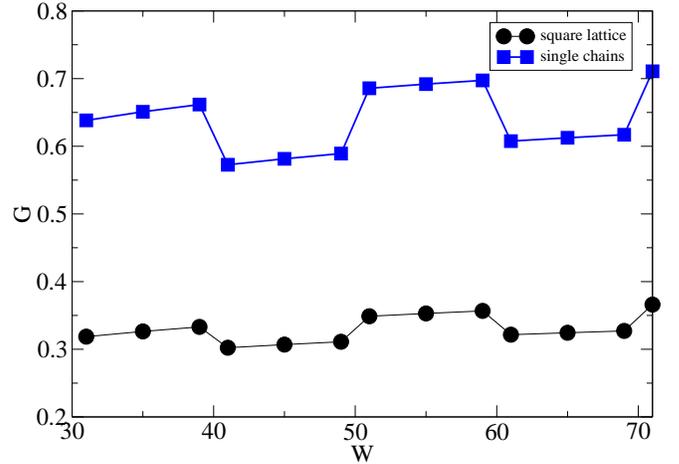}
\end{center}
\caption{The conductance $G$ for a pristine/clean($w=0$) nanowire vs its width $W$ for $L=400$, with two types of leads, square lattice and single chains.}
\label{fig4}
\end{figure}

\begin{figure}
\begin{center}
\includegraphics[width=1.0\columnwidth,clip=true]{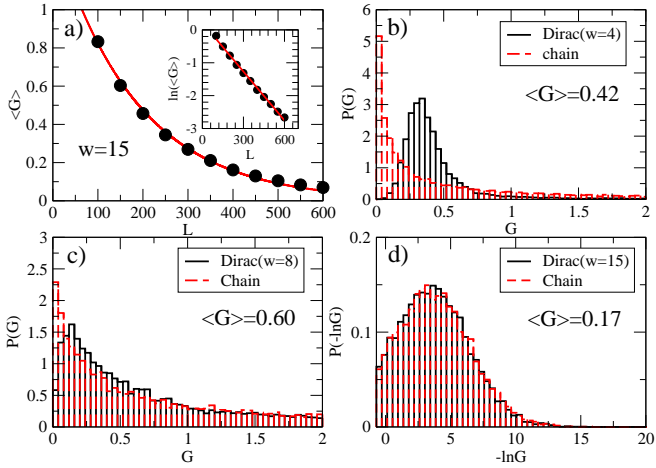}
\end{center}
\caption{The scaling of the conductance and its statistics for the 2D Dirac semimetal nanowire with width $W=61$ and even lengths $L$. For strong disorder $w=15$ in panel a), the average $\langle G \rangle$ decays exponentially with $L$ (solid red curve), as $\langle G \rangle \sim \exp (-\alpha L)$, which is equivalent to the linear behavior shown in the inset for the $\ln(\langle G \rangle)$, indicating Anderson localization.
The probability distributions of $G$ at various $w$ is plotted in panels b),c),d) showing the gradual transition toward the insulating phase. The red dashed histogram is derived from a disordered tight-binding chain, containing the standard Anderson localized (insulating) phase.}
\label{fig5}
\end{figure}

\begin{figure}
\begin{center}
\includegraphics[width=1.0\columnwidth,clip=true]{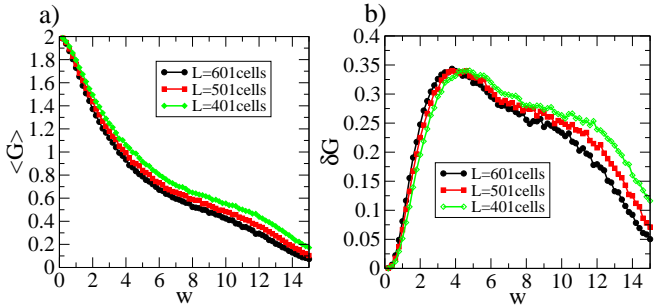}
\end{center}
\caption{The conductance statistics in the presence of ondiagonal disorder using the same parameters as in figure \ref{fig3}  but for odd $L$. The system behaves as a conventional disordered system lacking the topological phase for weak w unlike the nanowire with even $L$. For $w=0$ resonant scattering via two states at $E=0$ results in $G=2$, which decays with increasing w due to Anderson localization.}
\label{fig6}
\end{figure}


\begin{figure}
\begin{center}
\includegraphics[width=1.0\columnwidth,clip=true]{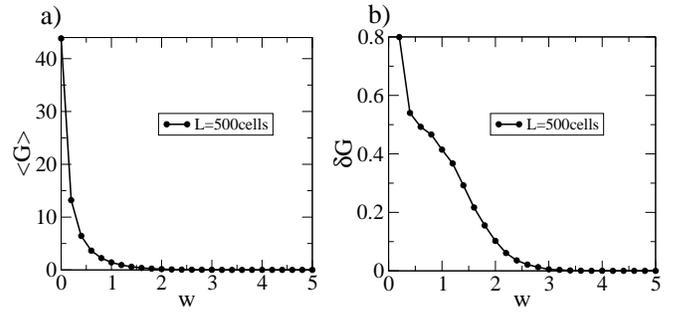}
\end{center}
\caption{The average conductance and its fluctuations for the tilted tight-binding square lattice without SOC, corresponding to $t_{so}=0eV$ with on-diagonal disorder for width $W=61$ and length $L=500$. The behaviors resemble that of the Dirac system for odd length $L$, with the difference being the large number of open channels for $w=0$.}
\label{fig7}
\end{figure}
\section{Non-Integer Conductance} 
In figure \ref{fig2}a, we plot the bandstructure of the 2D Dirac semimetal nanowire with $t_{so}=4.5eV$ for width $W=61$. The energy dispersion is linear near the Fermi energy ($E=0$), where the Dirac point lies, as shown in the inset of figure \ref{fig2}a. We provide analytical dispersion relations in the first section of the Appendix. We study the conductance statistics at this energy. We use two statistical measures, the average of the conductance $\langle G \rangle$ and its variance (fluctuations) $\delta G$ for an ensemble of 10000 realizations of the disorder. In figure \ref{fig3}, we plot $\langle G \rangle$ and $\delta G$ versus the disorder strength $w$ for an even number of unit cells in the nanowire (length $L$). For small values of the disorder strength ($w\lesssim2$) the average conductance $\langle G \rangle$ remains constant at an non-integer value $\langle G \rangle \approx 0.32$, irrespectively of the nanowire length $L$. Additionally the variance $\delta G$ is zero inside this area, indicating robustness to weak disorder as in other topological systems, such as the Kane-Mele(KM) and the Bernevig-Hughes-Zhang (BHZ) models\cite{paper3}. As we increase the disorder strength the system undergoes a transition/crossover towards the Anderson localized (insulating) phase, which is reached for strong disorder $w>10$. For the intermediate area ($ 2\lesssim w \lesssim 10 $), we observe an increase of $\langle G \rangle$ with w which does not usually occur in conventional disordered systems, where the wavefunctions become more localized with increasing disorder resulting in reduced $G$. The increase of $G$ with increasing disorder for the 2D Dirac semimetals, could be attributed to a mechanism of diffusion of the topological states for $w=0$ inside the bulk of the system as the disorder is increased. A similar mechanism can be observed in graphene nanostructures with zig-zag edges, which contain edge states, that are not topological however \cite{kleftogiannis,wang3}. Also, a similar effect due to contact and disorder scattering has been observed in bilayer graphene systems containing an energy gap \cite{xu}. Therefore by varying the disorder strength in the 2D Dirac semimetal we observe a transition/crossover from a topological phase, to a non-standard disordered phase and finally to an insulating phase. We note that usually the conductance in topological phases, such as the ones occurring in topological insulators, has a robust integer value, proportional to the number of open edge state channels present at $E=0$, resulting in $G=2$ for both the KM and the BHZ models\cite{paper3}. As we show for the 2D Dirac semimetal nanowire the conductance is not only robust to disorder but its value can be also non-integer. This effect can be attributed to the scattering/interference effects occuring at the interface between the leads and the nanowire(scatterer) which reduces $G$ to non-integer values, from the usual value $G=2$ occuring in topological insulators. Then this non-integer $G$ is stabilized inside the nanowire due to the topological properties of the 2D Dirac semimetal. In figure \ref{fig2}b we plot the energy levels of a nanowire sample for $L=30$ and $L=31$ and for the same width as in the transport calculations $W=61$. Despite the absence of states at $E=0$ for $L=30$, resulting in an energy gap, where the transport is performed, $G$ obtains the non-integer value that we have shown. An example of scattering through a dimer with a staggered potential, via two tight-binding chains as leads, demonstrating the transport mechanism through an energy gap, leading to the non-integer $G$, can be found in the first section of the Appendix. The non-integer values of $G$ in the 2D Dirac semimetal can be controlled by changing the width of the nanowire or the material used in the leads, which will modify the scattering effects at the interface between the leads and the nanowire. An example is shown in  figure \ref{fig4} where $G$ is plotted for pristine (clean) nanowires vs their width for two different types of leads, square lattice and single chains.
In figure \ref{fig5} we examine the scaling of $G$ and its probability distributions. For strong disorder $w=15$ shown in the figure \ref{fig5}a), the average $\langle G \rangle$ decays exponentially as $G \sim \exp (-\alpha L)$, with the length $L$ of the nanowire, as shown by the fitting red solid curve. This indicates that the system has reached the Anderson localized phase, with localization length $\frac{1}{\alpha}$, i.e. the 2D Dirac semimetal behaves similarly to conventional disordered systems for strong disorder.
A comparison between the probability distributions of
a disordered tight-binding chain with on-site disorder (after multiplying the conductance with two to simulate the spin degree of freedom) and the nanowire, are shown in figure \ref{fig5}b),c),d). The distributions from the two systems converge gradually as the disorder is increased. For $w=4$ the distribution of the nanowire spreads around the value $G \approx 0.32$ which is the robust non-integer value in the area $w\lesssim2$. This distribution resembles the ones observed at the crossover between the topological(QSH) and the metallic phase in the BHZ model at the presence of disorder\cite{paper3}. Then the distribution gradually approaches the one of the chain as w is increased. An almost perfect agreement can be seen for $w=15$, showing that the conductance statistics through the nanowire for strong disorder are identical to those occurring in 1D systems at the insulating regime.

We note that using the Dirac semimetal in the leads would restore the integer values of the the conductance, since the effect that we described is based on the material mismatch at the contacts(interface) between the leads and the nanowire(scatterer).

It would be tempting to relate the non-integer conductance values that we have found with the fractionalization effects
encountered in systems like the fractional-quantum-Hall-effect(FQHE)\cite{tsui,laughlin,stormer,wang} or the Kitaev chain\cite{kitaev,vernek}.
However, the non-integer values of $G$ for the 2D Dirac semimetals calculated, are irrational numbers and cannot be represented as fractions.

\section{Integer Conductance} 
In figure \ref{fig6}, we plot $\langle G \rangle$ and $\delta G$ versus the disorder strength $w$ for odd nanowire lengths $L$. In the absence of disorder the conductance has the integer value $G=2$. This is due to resonant scattering via two zero energy states, which can be seen in the energy spectrum in  figure \ref{fig2}b) for $L=31$. These two states create two open channels at the Fermi energy ($E=0$), for the quantum transport through the nanowire, via resonant scattering. We demonstrate a similar transport mechanism for an 1D system in the second section of the Appendix. Also, in the third section of the Appendix we present a chiral sublattice mechanism of the lattice structures of the Hamiltonian Eq. \ref{eq_1} that leads to the persistent states at $E=0$ for odd $L$. As seen in figure \ref{fig6} the mean conductance $\langle G \rangle$ is decreased with increasing  disorder, as the system undergoes a transition/crossover toward the insulating phase. Unlike the case with even $L$, there is no topological phase for weak disorder corresponding to $\delta G \sim 0 $, or an intermediate regime where $G$ is increased, before the onset to the insulating phase. A similar behavior of $G$ is also encountered in conventional 2D disordered systems, such as a square tight-binding lattice with on-site disorder. For example, in figure \ref{fig7} we show the statistics of $G$ for the titled square lattice after removing the SOC term, corresponding to $t_{so}=0eV$ in the 2D Dirac semimetal nanowire. The main differences with the behavior observed in figure \ref{fig6}, are the large number of open channels at $E=0$, the increased fluctuations of $G$ and a weaker stability to the disorder, due to the absence of SOC.

\section{Summary and conclusions}
We have studied numerically the statistics of the conductance G at the Dirac point of 2D Dirac semimetal nanowires, at the presence of short-range disorder. For an even number of cells in the nanowires (length $L$) we have found robust non-integer values for $G$ which remain unaffected by weak disorder and persist for any even $L$. The exact non-integer values of $G$ can be fine-tuned by the width of the nanowires and the material used in the leads to achieve the quantum transport through the nanowires. The effect can be attributed to the scattering effects at the interface between the leads and the nanowire, an energy gap for nanowires with even L and the topology of the 2D Dirac semimetals, which stabilizes the non-integer $G$ inside the nanowires. In addition we have studied the scaling of $G$ with $L$ and the probability distributions of $G$, showing how the system transitions from the topological phase for weak disorder to the Anderson localized(insulating) phase/regime for strong disorder, via a non-standard disordered phase. For odd $L$ the effect of robust non-integer $G$ disappears and the system behaves as a more conventional disordered system, starting with an integer $G$ at the absence of disorder and approaching gradually the insulating regime as the disorder strength is increased. We have found that this is due to resonant scattering via persistent states at the Dirac point where the quantum transport calculation is performed.

In short we have shown that it is possible to achieve robust non-integer quantum transport in 2D Dirac semimetal nanowires by controlling their geometry and the materials used in the quantum transport setup. This could have wider implications/connections on possible ways to achieve fractionalization effects in topological materials via quantum transport.

\section*{Acknowledgements}
We acknowledge support by the Project HPC-EUROPA3(INFRAIA-2016-1-730897), funded by the EC Research Innovation Action under the H2020 Programme. In particular, we gratefully acknowledge the computer resources and technical support provided by ARIS-GRNET and the hospitality of the Department of Physics at the University of Ioannina in Greece.

\section{Appendix}

\subsection{Analytical dispersion relations}

\begin{figure}
\begin{center}
\includegraphics[width=1.0\columnwidth,clip=true]{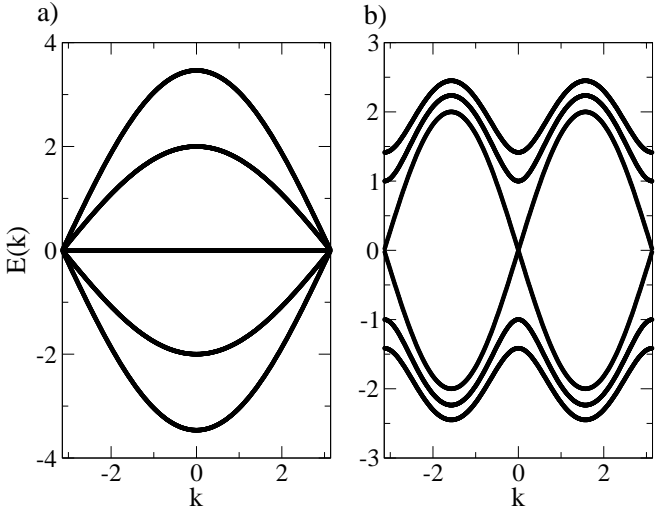}
\end{center}
\caption{a) The bandstructure of a rotated square tilted lattice using Eq. \ref{eq_disp1} in the Appendix for width  $N_{y} = 5$ and $t = 1eV$.  b) The bandstructure for the spin-orbit-coupling term using Eq. \ref{eq_disp2} in the Appendix with $t_{so}=1eV$.}
\label{fig8}
\end{figure}

The dispersion relation for a rotated square lattice nanoribbon (see figure \ref{fig1}), of width $W=N_y$, where $N_y$ is the number of sites in the unit cell, described by Eq. \ref{eq_2} with $t_{so}=0eV$, is
\begin{equation}
E(k,q) = 4 t cos\left(\frac{k}{2}\right) cos\left(\frac{\pi q}{N_{y}+1}\right).
\label{eq_disp1}
\end{equation}
Here $t$ is the hopping between neighboring sites, the variable $q$ takes the discrete values $q=1,2 ...N_{y}$, while $k$ is a continuous variable in the range $-\pi<k<\pi$, where the first Brillouin zone lies. One persistent feature of Eq. \ref{eq_disp1} is a flat band at $E=0$ for odd $N_y$. The bandstructure for $N_y=5$ and $t=1eV$ is shown in figure \ref{fig8}a).

The dispersion relation for the spin-orbit-coupling term in Eq. \ref{eq_2} with $t=0eV$,
consisting of two decoupled square lattice structures, between the same A-A or B-B sublattice sites in the rotated square lattice, is
\begin{equation}
E^{i}(k,q) = \pm t_{so}\sqrt{4 - 2cos\left(2k\right)- 2cos\left( \frac{\pi q_{so}}{N^{i}_{y^{'}}+1}\right)}.
\label{eq_disp2}
\end{equation}
The index $i$ denotes the sublattice A or B, $t_{so}$ is the SOC strength, $q_{so}$ is a variable taking discrete values and $N^{i}_{y^{'}}$ is the number of sites inside the unit cell of the nanowire, for each sublattice. If $N^{i}_{y^{'}}$ is even, then
\
\[ 
N^{i}_{y^{'}} = 
\begin{array}{ll}
      \frac{N_{y}}{2} & i=A,B  
\end{array} 
\]
while for $N^{i}_{y^{'}}$ odd, we have
\[ 
N^{i}_{y^{'}}= \left\{
\begin{array}{ll}
      \frac{N_{y}+1}{2}  & i = B \\
       \frac{N_{y}-1}{2}  & i = A.  
\end{array} 
\right. 
\]
The variable $q_{so}$ for each sublattice $i$=A,B takes the following values,
\[ 
 q_{so}= \left\{
\begin{array}{ll}
      1,3..,N^{i}_{y^{'}}  & \textrm{when } N^{i}_{y^{'}} \textrm{is even} \\
      0,2..,N^{i}_{y^{'}} & \textrm{when } N^{i}_{y^{'}} \textrm{is odd}
\end{array} 
\right. 
\]
We plot the bandstructure using Eq. \ref{eq_disp2} for $N_y=5$ and $t_{so}=1eV$ in figure \ref{fig8}b), where the Dirac point with a linear dispersion near $E=0$ can be seen. We note that Eq. \ref{eq_disp2} describes the form of the band structure due to the SOC term, despite ignoring the spin degree of freedom which results in doubly degenerate bands.

\subsection{Non-integer conductance}

\begin{figure}
\begin{center}
\includegraphics[width=1.0\columnwidth,clip=true]{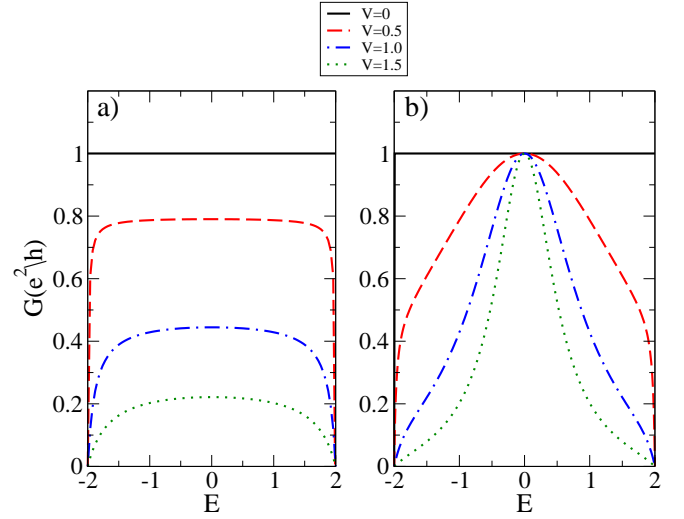}
\end{center}
\caption{a)The conductance $G$ of a dimer with potentials V,-V, with two tight-binding chains as leads. $G$ is non-integer, as for the 2D Dirac semimetal nanowires for even lengths $L$. b) $G$ for a chain with three sites with potentials V,0,-V demonstrating resonant scattering via one state at $E=0$, resulting in $G=1$, as for the nanowires of odd $L$.}
\label{fig9}
\end{figure}

In this section we derive analytically the conductance of a dimer consisting of two sites with staggered potential V and -V, by attaching two semi-infinite tight-binding chains at its ends acting as the leads. The Hamiltonian of the dimer chain is
\begin{equation}
H_{scat} =   \begin{bmatrix}  V & t \\ t & -V  \end{bmatrix},
\label{h_dim}
\end{equation}
where t is the hopping between the two sites.
The Hamiltonian Eq.\ref{h_dim} gives two states at energies $E=-\sqrt{t^2+V^2},\sqrt{t^2+V^2}$ and no states at $E=0$. The transmission probability through the dimer at any $E$ can be calculated via the Green's function
\begin{equation}
G(E) =  (EI-H_{scat}-\Sigma(E))^{-1}, 
\label{green}
\end{equation}
where I is the identity matrix and $\Sigma(E)$ is the self-energy matrix
\begin{equation}
\Sigma(E) =  \begin{bmatrix}  \Sigma_{L}(E) & 0 \\ 0 & \Sigma_{R}(E)  \end{bmatrix},
\label{semat}
\end{equation}
where $\Sigma_{L}(E), \Sigma_{R}(E)$ are the self-energies from the left and right leads. For tight-binding chains as leads which have energy dispersion $E=2t\cos{k}$, we have 
\begin{equation}
\Sigma_{L}(E)=\Sigma_{R}(E)=\frac{E}{2}-i\sqrt{1-\frac{E^2}{4}} for -2<E<2,
\label{se}
\end{equation}
for t=1. By plugging Eq. \ref{h_dim},\ref{semat},\ref{se} into Eq. \ref{green} and taking the appropriate Green's function element that is related to the transmission probability between the two sites in the dimer we have  
\begin{equation}
G_{21}(E)=\frac{2t}{E^{2} + i E\sqrt{4-E^{2}} -2(1+t^{2}+V^{2})}.
\label{green1}
\end{equation}
The actual transmission probability can be calculated via the Fisher-Lee relations
\begin{equation}
T_{12}(E)=v(E)^2|G_{21}(E)|^2,
\label{tr}
\end{equation}
where $v(E)=\frac{\theta E}{\theta k}=-2\sqrt{1-\frac{E^2}{4}}$ is the group velocity of the chains assuming that $\hbar=1$. Conductance is $G=\frac{e^2}{h}T_{12}$ which is plotted in figure \ref{fig9}a) for t=1 and various V. Despite the fact that the dimer does not contain states at E=0 a non-integer conductance manifests at this energy due to the scattering from the dimer of the incident electrons coming from the perfect metallic leads. A similar effect occurs for the transport through the 2D Dirac semimetal nanowire, the difference there being that the non-integer $G$ is stabilized inside the nanowire due to the topology of the system.
\subsection{Resonant transport}
In order to create resonant transport we consider
a scatterer consisting of a chain with three sites
and potentials V,0,-V respectively on each site.
The Hamiltonian of the system is
\begin{equation}
H_{scat} =  \begin{bmatrix}  V & t & 0 \\ t & 0 & t \\ 0 & t & -V \end{bmatrix}.
\label{h_trim}
\end{equation}
This system has two states at energies $E=-\sqrt{2t^2+V^2},\sqrt{2t^2+V^2}$ and one state at energy $E=0$. By following the same procedure as in the section above with
\begin{equation}
\Sigma(E) =  \begin{bmatrix}  \Sigma_{L}(E) & 0 & 0 \\ 0 & 0 & 0 \\ 0 & 0 & \Sigma_{R}(E) \end{bmatrix},
\label{semat1}
\end{equation}
we derive the transmission probability
\begin{equation}
T_{13}(E)=v(E)^2|G_{31}(E)|^2.
\label{tr1}
\end{equation}
The conductane $G=\frac{e^2}{h}T_{13}$ which is plotted in figure \ref{fig9}b) has an integer value at $E=0$ as for the 2D Dirac semimetal nanowires. The transport in both the chain and the nanowire are conducted via states that have the same energy as the incident electrons from the leads causing resonant transport with $G=1$ for the chain and $G=2$ for the nanowire.

\subsection{Chiral/Bipartite Mechanism}

\begin{figure}
\begin{center}
\includegraphics[width=1.0\columnwidth,clip=true]{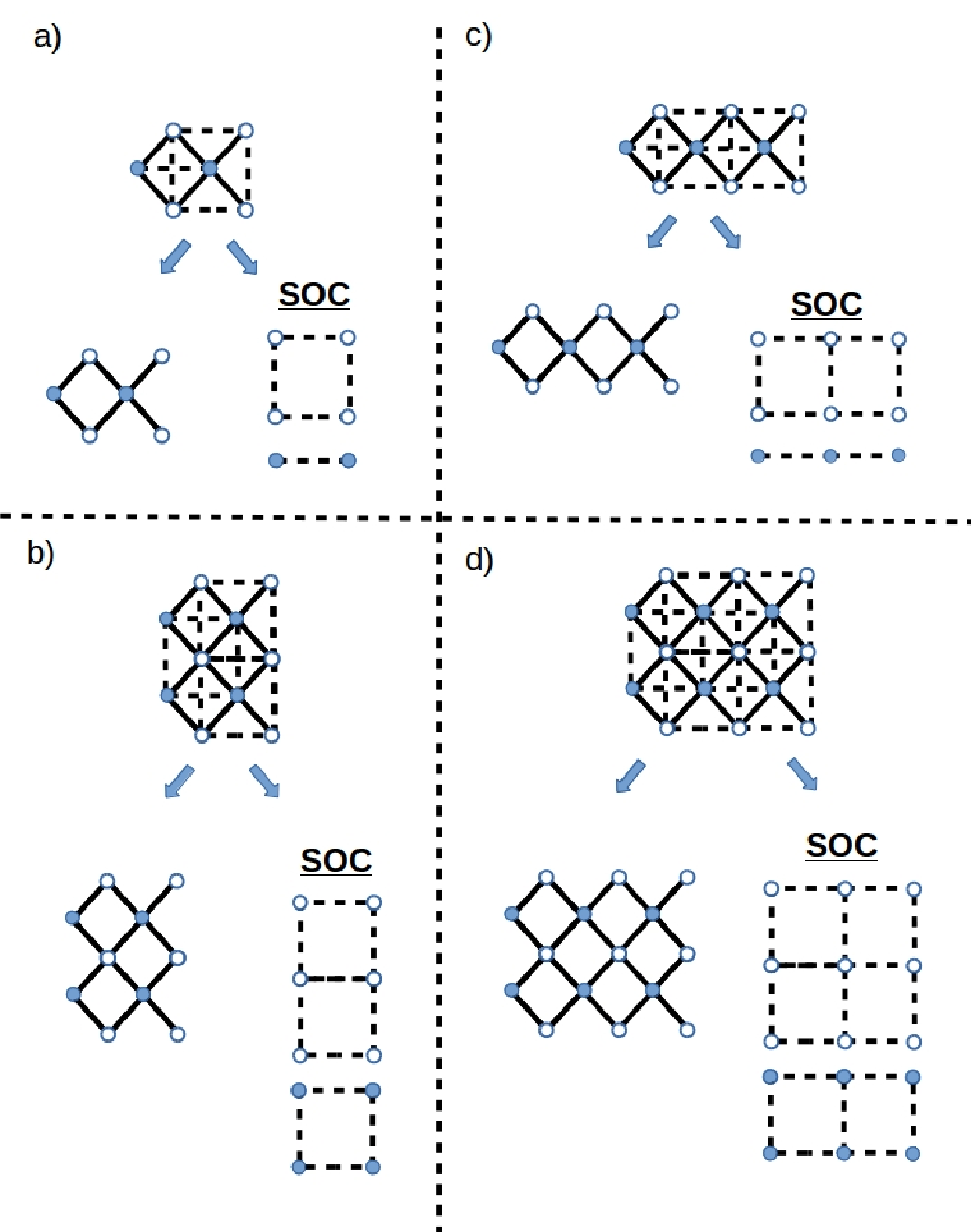}
\end{center}
\caption{The tight-binding lattice used to model the Dirac semimetal nanowire for even and odd lengths $L=2,3$ and two different widths $W=3,5$. The lattice can always be decomposed into two terms, the tilted square lattice
consisting of two interconnected sublattices(filled(A) and empty(B) sites) and two regular lattices describing the SOC term between sublattice sites of the same type A or B in the tilted lattice. For an odd number of sites in either of the SOC lattices an $E=0$ state appears in the energy spectrum of the nanowire, due to a chiral/bipartite lattice symmetry.}
\label{fig10}
\end{figure}

The appearance of the $E=0$ states for an odd number of unit cells (odd length $L$) in the Dirac semimetal nanowire can be understood by examining the contribution of the different lattice structures in the Hamiltonian Eq. \ref{eq_1}. A few examples can be seen in figure \ref{fig10} for nanowires with width $W=3,5$ and an even($L=2$) (a),b)) and odd ($L=3$) (c),d)) number of unit cells. In all cases the system can be  decomposed into a tilted square lattice shown on the left side of the individual panels and another lattice describing the SOC interaction on the right side. Filled(empty) circles denote A(B) sublattice sites. The SOC interaction connects sites of the same sublattice giving two SOC lattices one for the A and one for the B sublattice.  The tilted square lattice contains $N_{B}>N_{A}$ sites and is bipartite containing a chiral symmetry between the A and B sublattices\cite{inui}. This symmetry leads to $N_{B}-N_{A}$ states at $E=0$. These states are responsible for the $E=0$ flat band in the band structure shown in figure \ref{fig8}a) for odd $N_y$(nanoribbon width). The same argument can be applied for the SOC lattices, which can be splitted also in two sublattices say C and D. Whenever one of these SOC lattices contains an odd number of total sites it gives at least one state at $E=0$, since $N_{C}-N_{D}=1$\cite{inui} (taking also account of the spin degree of freedom we get two $E=0$ states).  For example in figure \ref{fig10}c) the bottom SOC lattice connecting A sites is a chain of three sites, while in figure \ref{fig10}d) the upper SOC lattice has nine total sites. Notice that the $E=0$ states are absent for an even nanowire length $L$, since the corresponding SOC lattices always have an even total number of sites, as shown in the examples in figures \ref{fig10}a),b). Since $t_{so}>t$ in Eq. \ref{eq_1} the chiral mechansim of the SOC lattices dominates the behavior of the system at $E=0$. To summarise two states appear for the Dirac semimetal nanoribbon for odd number of unit cells, at the band center, at energy $E=0$, due to a chiral sublattice symmetry of the SOC lattice structures. The $E=0$ states lead to the resonant scattering transport mechanism that we have shown for odd nanowire lengths.

\end{document}